\documentstyle[aps,twocolumn,epsf]{revtex}

\begin{document}

\title{Fluorescence decay in aperiodic Frenkel lattices}

\author{Francisco Dom\'{\i}nguez-Adame$^{\dag}$ and Enrique
Maci\'{a}$^*$}

\address{Departamento de F\'{\i}sica de Materiales, Facultad de
F\'{\i}sicas, Universidad Complutense, E-28040 Madrid, Spain \\
\bigskip}

\author{\small\parbox{14cm}{\small
We study motion and capture of excitons in self-similar linear
systems in which interstitial traps are arranged according to an
aperiodic sequence, focusing our attention on Fibonacci and Thue-Morse
systems as canonical examples.  The decay of the fluorescence intensity
following a broadband pulse excitation is evaluated by solving the
microscopic equations of motion of the Frenkel exciton problem.  We find
that the average decay is exponential and depends only on the
concentration of traps and the trapping rate.  In addition, we observe
small-amplitude oscillations coming from the coupling between the
low-lying mode and a few high-lying modes through the topology of the
lattice.  These oscillations are characteristic of each particular
arrangement of traps and they are directly related to the Fourier
transform of the underlying lattice.  Our predictions can be then used
to determine experimentally the ordering of traps.
\\[3pt]PACS numbers: 71.35+z; 36.20.Kd; 78.90.+t}
}\address{}\maketitle

\section{Introduction}

The discovery of quasicrystals \cite{Shechtman} and the fabrication of
aperiodic low-dimensional systems \cite{Merlin} have encouraged the
study of the physical properties of solids with long-range order,
lacking translational symmetry.  Physical aperiodic systems, whose
structural order is described by means of deterministic sequences,
present rather exotic electronic properties not shared by crystalline
and amorphous solids.  Peculiar signatures of these unique electronic
states are Cantor-like spectra and self-similar wave functions.  Several
electronic properties of solids can be inferred from optical
characterization techniques like optical absorption, photoluminescence
and fluorescence after pulse excitation.  Therefore, a complete
understanding of the interplay between the electronic properties and the
underlying aperiodic order requires a detailed analysis of the optical
dynamics of this kind of ordering of matter.  This interest has
motivated various works dealing with optical properties of aperiodic
systems, mainly devoted to Fibonacci and Thue-Morse semiconductor
superlattices. \cite{DasSarma,SSC,Tuet,Munzar}

{}Following a long term project regarding electronic and transport
properties of aperiodic systems, \cite{PLA,PRBFIBO,PRE,SST} our group
focused its attention on optical absorption spectra of Frenkel excitons
in Fibonacci and Thue-Morse lattices. \cite{OPTFIBO,masterfib} Our main
aim was to learn about the phenomenology of incoherent and coherent
exciton dynamics in molecular aggregates and polymers exhibiting
long-range correlations.  We considered Fibonacci and Thue-Morse
lattices as canonical aperiodic models that are neither periodic nor
random.  In turn, they are inherently close to realistic aperiodic
systems where the effects of long-range order might change dramatically
the exciton dynamics.  We found numerically several characteristic lines
specific of each aperiodic system that are not present in the spectra of
periodic or random Frenkel lattices, thus being an adequate way for
determining the particular structural order of the system from
experiments.\cite{OPTFIBO} Moreover, using a model based on the Pauli
master equation, we have also studied motion and capture of incoherent
excitons when traps are arranged according to aperiodic sequences.
\cite{masterfib} As an interesting feature, amenable to experimental
confirmation by means of luminescence studies at moderate temperature,
we obtained that the decay of the survival fraction of incoherent
excitons is simply exponential $P(t)\sim \exp(-At)$ instead of the
asymptotic stretched exponential $P(t)\sim \exp(-At^{1/2})$ appearing in
one-dimensional random lattices. \cite{masterran} It is well-known that
the exponent of the stretched exponential varies when passing from
trapping of classical (incoherent) to quantum (coherent) excitons in
random lattices. \cite{Parris} Then a natural question arises in the
context of aperiodic systems, namely whether the simple exponential
behavior of the survival fraction holds in the quantum case.

In this work we investigate the decay fluorescence due to Frenkel
excitons in linear systems with interstitial traps arranged according to
the Fibonacci and Thue-Morse sequences.  We have focused on the analysis
of aperiodic arrangements of traps in systems where optically active
centers are placed in a straight line.  Nevertheless, the model can also
be applied to other possible geometric configurations of interest as,
for example, zig-zag chains whose $sp^2$ hibridization is explicitly
considered.  To this end, we make use of a general treatment which
allows us to study the dynamics of Frenkel excitons in these lattices,
solve the microscopic equations of motion and find the fluorescence
intensity as a function of time.  The main results of this paper are the
following.  First, we find out the decay law of the fluorescence
intensity at low temperature, and compare it to the above mentioned
simple exponential trend at moderate temperature.  Second, and most
important from an experimental point of view, we obtain {\em distinctive
features} of the decay curves $P(t)$ allowing us to unveil the ordering
of traps present in the system.  By means of an analytical approach we
are able to explain the origin of these characteristics oscillations,
which are related to the Fourier transform of the arrangement of traps.
Thus we successfully relate an optical property (fluorescence decay)
with an structural one (topology of the lattice).

The remaining of the paper is organized as follows.  In Sec.~II we
describe our model Hamiltonian and the way we arrange the traps
following the Fibonacci and Thue-Morse sequences.  Afterwards, in
Sec.~III we focus our attention on the fluorescence intensity following
a broadband pulse excitation and solve the microscopic equation of
motion.  We find numerically that, superimposed to an overall
exponential decay, there exist small-amplitude oscillations
characteristic of each particular system.  In Sec.~IV we develop a
homogeneous medium approximation which accounts for the exponential
decay, and, in Sect.~V we introduce a perturbative treatment that
successfully explains the small-amplitude oscillations.  In this way, we
clearly demonstrate that the frequency of the oscillations depends on
the specific aperiodic order of the distribution of traps and, in
addition, that these oscillations arise as a consequence of the coupling
between the low-lying mode and few high-lying modes through the topology
of the lattice.  Section~VI concludes the paper with a brief summary of
results and some general remarks on their physical implications.

\section{Physical Model and Theory}

\subsection{Model Hamiltonian}

We consider a system of $N$ optically active two-level centers,
occupying positions ${\bf r}_n$ on a linear regular lattice with spacing
$a$.  For our present purposes we neglect all thermal degrees of
freedom.  Therefore, the effective Frenkel Hamiltonian describing this
system can be written (we use units such that $\hbar=1$)
\cite{Parris,Parris89,Huber90,Wager}
\begin{equation}
{\cal H}= \sum_{n}\> (V_{n}-i\Gamma_n) a_{n}^{\dag}a_{n} +
\sum_{l\neq n}\>J_{nl}a_{n}^{\dag}a_{l}.
\label{Hamiltonian}
\end{equation}
Here $a_{n}^{\dag}$ and $a_{n}$ are Bose operators creating and
annihilating an electronic excitation of energy $V_n$ at site $n$,
respectively. $J_{nl}$ ($n\neq l$) is the intersite interaction of
dipole origin between centers $n$ and $l$.  In the case where all
centers have equal and parallel dipole moment the interaction can be
cast in the form $J_{nl}=-J(a/|{\bf r}_n-{\bf r}_l|)^3$, where $J$ is
the coupling between nearest-neighbor centers.  In this paper $J$ is the
unit of energy and $J^{-1}$ the unit of time.  Moreover, since $J_{nl}$
is a rapidly decreasing function of the distance between centers, we
will omit interactions beyond nearest-neighbors.

According to previous works, \cite{Parris89,Huber90,Wager,Frenkeltrap}
the non-Hermitian term $-i\sum_{n}\>\Gamma_{n}a_{n}^{\dag}a_{n}$
accounts for the irreversible trapping of the exciton due to
interstitial traps and the subsequent photon emission.  Here $\Gamma_n =
\Gamma$ if the $n$ center has a trap associated with it and $\Gamma_n =
0$ otherwise, $\Gamma$ being the trapping rate.  The site energy $V_n$
can be written, without loss of generality, as $\langle V \rangle+\Delta
V_n$.  The first term stands for the average excitation energy and we
drop it by setting an appropriate reference.  The second term is a
random variable reflecting the offset energy due to the influence of the
surrounding medium.  Since we are mainly interested in the trapping
dynamics, we neglect the effects of diagonal disorder ($\Delta V_n =
0$), noting that they could be easily incorporated in our treatment if
necessary.

\subsection{Fibonacci lattice}

In this work we will be concerned with the Frenkel exciton dynamics and
trapping processes in systems presenting an aperiodic distribution of
interstitial traps.  The Fibonacci lattice (FL) is the archetypal
example of deterministic and quasiperiodically ordered structure.  Any
arbitrary Fibonacci system presents two kind of building blocks.  In our
case, we choose those blocks as individual centers with and without a
trap associated with it, hereafter called $T$ and $S$, respectively.
The Fibonacci arrangement can be generated by the substitution rule
$S\rightarrow ST$, $T\rightarrow S$. In this way, finite and
self-similar aperiodic lattices are obtained by $n$ successive
applications of the substitution rule.  The {\em n\/}th generation
lattice will have $N=F_n$ elements, where $F_n$ denotes a Fibonacci
number.  Such numbers are generated from the recurrence law
$F_n=F_{n-1}+F_{n-2}$ starting with $F_0=F_1=1$; as $n$ increases the
ratio $F_{n-1}/F_n$ converges toward $\tau=(\sqrt{5}-1)/2=0.618\ldots$,
an irrational number which is known as the inverse golden mean.
Therefore, lattice sites are arranged according to the sequence
$S\,T\,S\,S\,T\,S\,T\,S\ldots$ This type of distribution of traps will
be referred to as {\em minority} lattice since the fraction of
$T$-centers ($1-\tau$ for large enough FLs) is smaller than the fraction
of $S$-center ($\tau$ for large enough FLs).  Moreover, it is worth
noticing that $T$-centers appear isolated in the minority lattice.  In
the same way, one can generate the {\em majority} lattice by replacing
$S \leftrightarrow T$, thus obtaining the sequence
$T\,S\,T\,T\,S\,T\,S\,T\ldots$ In this case $T$-centers can appear
isolated or paired.

\subsection{Thue-Morse lattice}

The Thue-Morse lattice (TML) is also constructed from two building
blocks $S$ and $T$.  The substitution rule in this case is $S\rightarrow
ST$, $T\rightarrow TS$ and the {\em n\/}th generation lattice contains
$2^{n}$ centers.  It is clear that the fraction of $T$-centers is
exactly $1/2$ for any generation lattice.  The resulting sequence
$S\,T\,T\,S\,T\,S\,S\,T\ldots$ is also aperiodic and self-similar but
not quasiperiodic.  Notice that both $S$- and $T$-centers appear
isolated or paired.  The TML present an inversion center so that the
lattice is invariant under the transformation $S \leftrightarrow T$.

\subsection{Computation of physical magnitudes}

Having presented our model we now briefly describe the method we have
used to calculate the fluorescence decay following a short-pulse
excitation.  The calculation involves the total dipole moment operator
${\cal D}=\sum_{n}\>d_{n}(a_{n}^{\dag}+a_{n}),$ where $d_{n}$ is the
dipole moment of the $n$ center.  Notice that we are restricting
ourselves to the case of systems whose length is much smaller than the
optical wavelength.  After the excitation, the state vector at time $t$
evolves according to $|\psi(t)\rangle = \exp(-i{\cal H}t){\cal
D}|\mbox{vac}\rangle$, $|\mbox{vac}\rangle$ being the exciton vacuum
state.  Besides radiative effects, the fluorescence intensity,
normalized to the value at $t=0$, is given by \cite{Huber90}
\begin{equation}
P(t)=\frac{|\langle\psi(0)|\psi(t)\rangle|^2}
          {|\langle\psi(0)|\psi(0)\rangle|^2}.
\label{fluor1}
\end{equation}
The fluorescence intensity can also be viewed as the probability of
finding an exciton in the ${\bf k}={\bf 0}$ mode at time $t$.

A reliable method for determining $P(t)$ numerically has been given by
Huber and Ching.\cite{Huber90} These authors introduced a set of
correlation functions
\begin{equation}
G_{n}(t)=\sum_{l}\>d_l\langle\mbox{vac}| a_{n}(t)a_{l}^{\dag}|\mbox{vac}\rangle,
\label{G}
\end{equation}
where $a_{n}(t)= \exp (i{\cal H}t) a_{n} \exp (-i{\cal H}t)$ is the
annihilation operator in the Heisenberg representation.  The function
$G_{n}(t)$ obeys the equation of motion
\begin{equation}
i{dG_{n}(t)\over dt} = \sum_{l}\> H_{nl} G_{l}(t),
\label{motion}
\end{equation}
with the initial condition $G_{n}(0)=d_{n}$.  The diagonal elements of
the tridiagonal matrix $H_{nl}$ are $-i\Gamma_{n}$ whereas off-diagonal
elements are simply given by $-J$.  The microscopic equation of motion
is a discrete Schr\"odinger-like equation on a lattice and standard
numerical techniques may be applied to obtain the solution.  Once these
equations of motion are solved, the fluorescence intensity is evaluated
from the relationship
\begin{equation}
P(t)=\frac{\left| \sum_{n}\> d_{n}^* G_{n}(t) \right|^2}
		  {\left| \sum_{n}\> |d_{n}|^2 \right|^2}.
\label{fluor2}
\end{equation}

\section{Numerical Results}

We have solved numerically the equation of motion (\ref{motion}) using
an implicit (Crank-Nicholson) integration scheme.\cite{Crank} We note
that energy is measured in units of $J$ and time in units of $J^{-1}$
and, since $4J$ is the exciton bandwidth, the energy and time scales can
be deduced from each particular experimental situation.  Aperiodic
lattices are generated using the inflation rules discussed above.  In
order to minimize end effects, spatial periodic boundary conditions are
introduced in all cases.  The maximum integration time and the
integration time step were $20$ and $2\times 10^{-3}$, respectively.
Smaller time steps give essentially the same results.  In most cases of
practical interest, one finds that the trapping rate is much less than
the exciton bandwidth, that is, $\Gamma \ll 4J$.  Thus, as typical
values, we have studied lattices with $\Gamma/J=0.01,\ldots,0.05$.  The
maximum lattice size under consideration was $N=F_{12}=233$ for FLs and
$N=2^{8}=256$ for TMLs.  Finally, local dipole moments are taken to be
$d_{n}=1/\sqrt{N}$.

\begin{figure}

\centerline{\epsfxsize=6.0truecm \epsffile{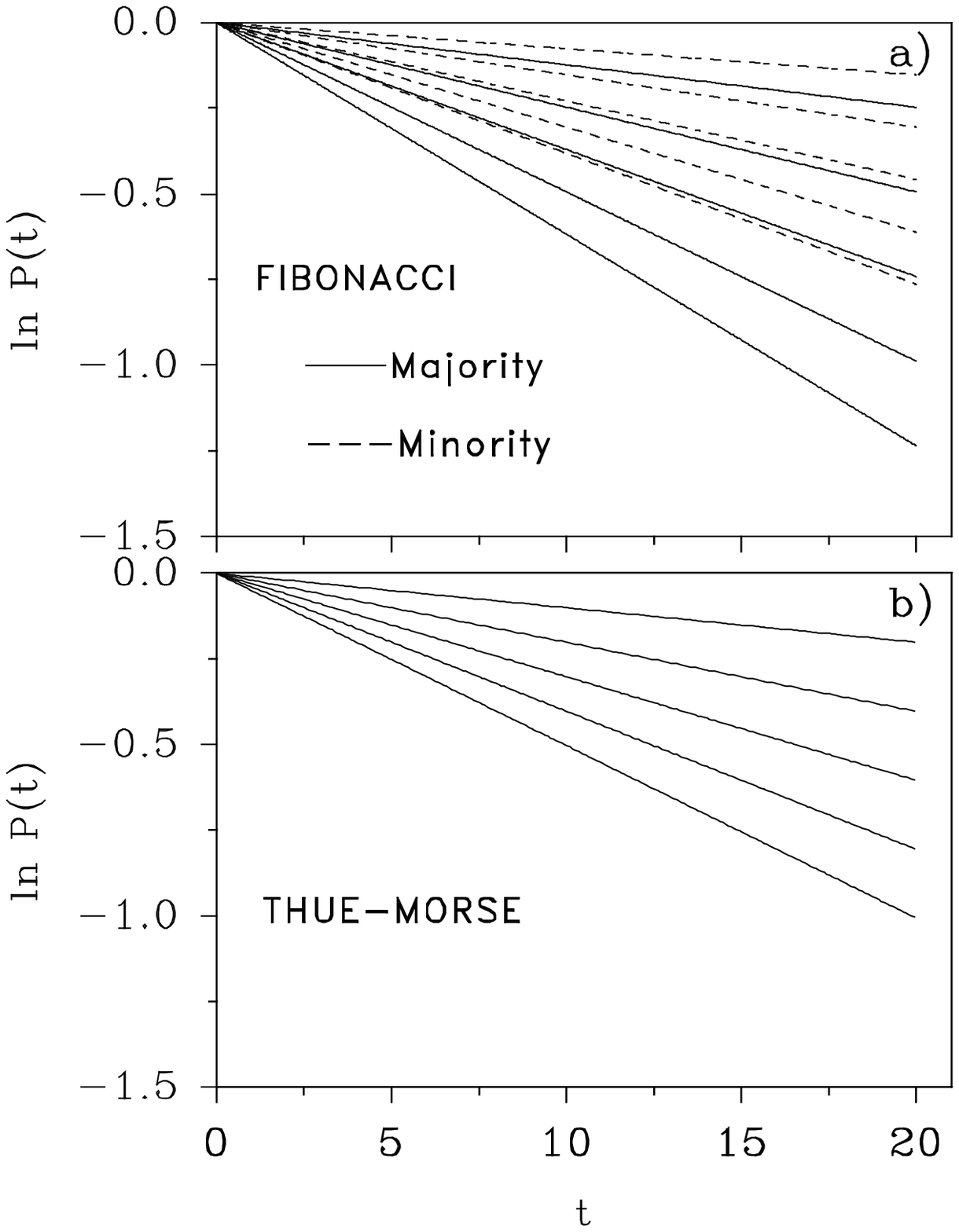}}\vspace{-1cm}

\caption{Time decay of fluorescence intensity in (a) Fibonacci lattices
with $N=F_{12}=233$ centers and a fraction of traps $c=0.618\ldots$
(majority lattice, solid lines) and $c=0.382\ldots$ (minority lattice,
dashed lines), and (b) Thue-Morse lattices with $N=2^{8}=256$ centers.
In all cases the trapping rate is, from top to bottom, $\Gamma =0.01,
\ldots, 0.05$ in units of $J$.}

\label{fig1}

\end{figure}

In our numerical simulations we have found that the fluorescence
intensity decays exponentially, for both FLs and TMLs, in the full time
interval considered by us.  Figure~\ref{fig1} shows the results for
majority and minority FLs with $N=233$ and TMLs with $N=256$, but we
have checked that this behavior is independent of the system size.  As
expected, the larger the trapping rate, the faster the decay of the
fluorescence intensity.  Moreover, for the same value of the trapping
rate, this decay is faster on increasing the fraction of traps (minority
FLs, TMLs and majority FLs).  Hence, as occurs in the case of completely
incoherent excitons, \cite{Frenkeltrap} the decay of Frenkel excitons in
FLs and TMLs resembles that of periodic lattices instead of random ones,
where the decay is stretched exponential.  The average decay of the
fluorescence intensity fits very well to an exponential of the form
$\exp(-2c\Gamma t)$ in the considered time interval, $c$ being the
fraction of traps in each lattice. This time dependence has also been
confirmed by fitting the data to stretched exponentials of the form
$\exp(-At^{\beta})$, obtaining in all cases $\beta = 1.0000 \pm
10^{-4}$.

\begin{figure}

\centerline{\epsfxsize=6.0truecm \epsffile{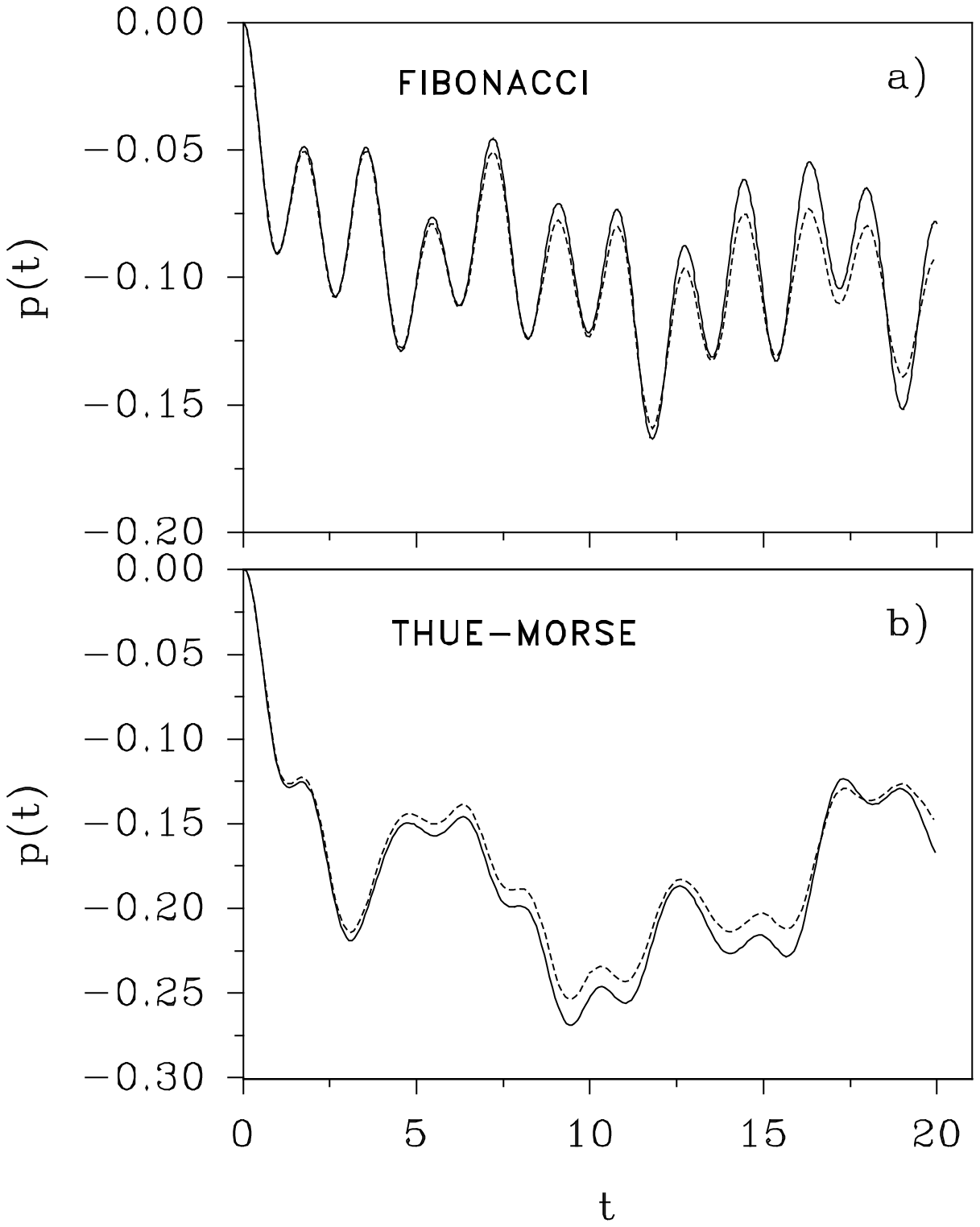}}\vspace{-1cm}

\caption{Plot of $p(t) \simeq \Gamma^{-2}\ln[ P(t)\exp(2c\Gamma t)]$ as
a function of time for (a) Fibonacci and (b) Thue-Morse lattice.  Solid
and dashed lines indicate numerical results and perturative prediction,
respectively.  Results are independent of the number of lattice sites
and the concentration of traps.}

\label{fig2}

\end{figure}

A detailed analysis of data plotted in Fig.~\ref{fig1} shows the
occurrence of small-amplitude oscillations superimposed to the average
exponential decay of the fluorescence intensity.  The functional form
$P(t)$ in all cases considered can be expressed as
\begin{equation}
P(t)=\exp(-2c\Gamma t)\left[ 1+\Gamma^2 p(t)\right],
\label{pdet}
\end{equation}
where $p(t)$ is an oscillatory function which is independent of the
trapping rate.  This function can be readily extracted from numerical
data as $p(t) \simeq \Gamma^{-2}\ln[ P(t)\exp(2c\Gamma t)]$.
Figure~\ref{fig2} shows the typical form of the $p(t)$ function for
anyone of the trapping rates $\Gamma$ considered in our study.
Remarkably, this function is, on the one side, independent of the number
of active centers present in the lattice and, most interestingly, it
does not depends on the concentration of traps, i.e. it is exactly the
same for both majority and minority versions of FLs ---this invariance
is, of course, trivial for TMLs---.  Finally, by comparing
Figs.~\ref{fig2}(a) and (b) we realize that the oscillatory pattern of
$p(t)$ is characteristic of the arrangement of traps, in the sense that
its general features are very sensitive to the kind of aperiodic order
being considered.

\section{Homogeneous medium approximation}

In this section we develop a theoretical approach to explain the overall
exponential decay of the fluorescence in aperiodic lattices.  To this
end we rewrite the system Hamiltonian ${\cal H}={\cal H}_{\text{HM}}
-i\Gamma{\cal H}_{\text{T}}$, where
\begin{mathletters}
\begin{eqnarray}
{\cal H}_{\text{HM}} &=& -ic\Gamma\sum_{n}\>a_{n}^{\dag}a_{n}
-J\sum_{n} \> (a_{n}^{\dag}a_{n+1} + a_{n+1}^{\dag}a_{n}),
\label{MF1} \\
{\cal H}_{\text{T}} &=& \sum_{n} \> (\Gamma_n/\Gamma-c)
a_{n}^{\dag}a_{n}.
\label{MF2}
\end{eqnarray}
\end{mathletters}
As a first approximation, we neglect the topology of the lattice and
take ${\cal H} \simeq {\cal H}_{\text{HM}}$.  Physically this implies
replacing the aperiodic distribution of traps by a uniform distribution
with a renormalized trapping rate $c\Gamma$. To proceed we notice that
the overlap between the state vector at time $t$ and the initial state
vector may be calculated from \cite{Huber92}
\begin{equation}
\langle\psi(0)|\psi(t)\rangle=\sum_{k}\>|\langle k | {\cal D} |
{\mbox{vac}}\rangle |^2 \exp(-iE_{k}t),
\label{overlap}
\end{equation}
where $|k\rangle$ denotes the one-exciton eigenvectors of (\ref{MF1})
and $E_k$ the corresponding eigenvalues. The Hamiltonian ${\cal
H}_{\text{HM}}$ with periodic boundary conditions can be exactly
diagonalized, yielding the eigenvectors
\cite{Fidder}
\begin{mathletters}
\label{dipole}
\begin{equation}
|k\rangle=\left({1\over N}\right)^{1/2} \sum_{n}
\exp \left( 2\pi i{nk\over N} \right) a_{n}^{\dag}|\mbox{vac}\rangle,
\label{vectors}
\end{equation}
with $k=0,\ldots,N-1$. The corresponding
eigenvalues read
\begin{equation}
E_{k}=-ic\Gamma-2J\cos\left(2\pi{ k\over N}\right)
\equiv -ic\Gamma+\omega_k.
\label{values}
\end{equation}
\end{mathletters}
Therefore, the dipole moment matrix elements in the main field approach
satisfy the relation $\langle k|{\cal D}|\mbox{vac}\rangle
=\delta_{k0}$.
Consequently one obtains that $\langle\psi(0)|\psi(t)\rangle=
\exp(-c\Gamma t-i\omega_{k}t)$ so that we obtain the exponential
behavior $P_{\text{HM}}(t)=\exp(-2c\Gamma t)$.  This result shows that,
within the tight-binding approximation, the overall exponential decay of
the fluorescence is a general property of {\em ordered} systems
described in terms of the Hamiltonian (\ref{Hamiltonian}), with
independence of this order being periodic or aperiodic.  We have
confirmed this point by considering periodic arrangements of traps with
different basis.  In all cases we have observed that the exponential
behavior predicted by the homogeneous medium approximation treatment
successfully describes the overall decay obtained by numerical
simulations.  On the other hand, the exact result obtained within the
that approximation provides us with a ground starting point which
will allow us to explain in full detail the small-amplitude oscillations
shown in Fig.~\ref{fig2}.

\section{Perturbative approximation}

The replacement of a given aperiodic distribution of traps by a uniform
one with a renormalized trapping rate is a crude representation of the
real system.  To develop a better approximation one must explicitly
consider the effects of the topology on the exciton decay through the
term $-i\Gamma{\cal H}_{\text T}$. To carry out such an approximation we
have inspired in the nondegenerate perturbation theory developed by
Huber to study {\em random} systems with substitutional traps.
\cite{Huber92}  Let $|\tilde{k}\rangle$ the perturbed eigenvector
describing the exciton state when the influence of the term ${\cal
H}_{\text T}$ is taken into account.  To evaluate the perturbed dipole
moment matrix element, we expand $|\tilde{k}\rangle$ in the basis of the
unperturbed eigenvectors $|k\rangle$ given in (\ref{vectors})
\begin{mathletters}
\begin{equation}
|\tilde{k}\rangle=|k\rangle-i\Gamma \sum_{l\neq k} {\langle l |{\cal
H}_{\text T}| k \rangle \over E_{k}-E_{l}}|l\rangle,
\label{pervector}
\end{equation}
whereas the perturbed eigenvalues are given by
\begin{equation}
\tilde{E}_k=E_{k}-i\Gamma\langle k |{\cal H}_{\text T}| k\rangle,
\label{pervalue}
\end{equation}
\end{mathletters}
with $E_k$ being the unperturbed eigenvalues.  Making use of the fact
that the dipole moment matrix element involving unperturbed eigenvectors
is $\langle k|{\cal D}|\mbox{vac}\rangle=\delta_{k0}$, the overlap
between $|\psi(0)\rangle$ and $|\psi(t)\rangle$ can be readily found,
leading to
\begin{eqnarray}
P(t)&=&P_{\text{HM}}(t) \left| {1+\Gamma^2 z(t)\over 1+\Gamma^2 z(0)}
\right|^2 \nonumber \\ &=& P_{\text{HM}}(t) \left\{ 1+2\Gamma^2\mbox{Re}
[z(t)-z(0)]\right\}+{\cal O}\left(\Gamma^4\right),
\label{rollo}
\end{eqnarray}
with
\begin{equation} z(t)\equiv\sum_{k\neq 0} {|\langle 0 |
{\cal H}_{\text T}| k \rangle |^{2} \over \omega_{k0}^{2}}
\exp(-i\omega_{k0}t),
\label{z}
\end{equation}
where we have defined $\omega_{k0} \equiv \omega_{k}-\omega_{0}=4J
\sin^{2}(\pi k/N)$.  To obtain the above result we have made use of the
result $\langle k |{\cal H}_{\text T} | k \rangle=0$, which according to
(\ref{pervalue}), indicates that $\tilde{E}_k =E_{k}$ in the first order
perturbative treatment.  Hence, recalling Eq.~(\ref{pdet}), one gets
\begin{equation} p(t)= 2 \sum_{k\neq
0}{|\langle 0|{\cal H}_{\text T}| k \rangle|^{2} \over \omega^{2}_{k0}}
\left(\cos\omega_{k0}t-1\right).
\label{antefinal}
\end{equation}
The most remarkable fact of this expression is that it relates the
small-amplitude oscillations of the fluorescence decay, an
experimentally measurable magnitude, with the Fourier transform of the
distribution of traps describing the topological ordering of the system.
To demonstrate this point, we recall that unperturbed eigenvectors are
orthogonal.  Thus
\begin{eqnarray}
\langle 0|{\cal H}_{\text T}| k \rangle &=& \langle 0 |\sum_{n} \>
{\Gamma_n\over\Gamma} a_{n}^{\dag}a_{n}| k \rangle \nonumber \\ &=&
{1\over N} \sum_{n} {\Gamma_n\over\Gamma} \exp\left( 2\pi i {kn\over
N} \right), \ \ \ \ \ \ k\neq 0.
\label{final}
\end{eqnarray}
which is nothing but the afore mentioned Fourier transform.  A
comparison of the perturbative prediction (\ref{antefinal}) with the
numerical result is given in Fig.~\ref{fig2}, where an excellent
agreement is achieved for both Fibonacci and Thue-Morse aperiodic
arrangements.  At this point we should stress that the above result is
valid for any arbitrary lattice.  It relates the Fourier pattern of the
distribution of traps with the finer details of the fluorescence decay.
Since each particular arrangement of traps leads to a different Fourier
pattern of $p(t)$, we arrive at the conclusion that the small-amplitude
oscillations characterize every distribution of traps in much the same
way as the usual X-ray diffraction analysis characterize the
distribution of scattering centers in the space.  It is also worth
mentioning that it is not difficult to show from (\ref{final}) that {\em
dual} lattices, namely those lattices obtained from the replacement $T
\leftrightarrow S$, present the same value of $|\langle 0|{\cal
H}_{\text T}|k\rangle|^{2}$.  In particular, minority and majority
Fibonacci lattices should display exactly the same oscillations, as it
is indeed the case in our numerical results.

In order to get a deeper insight into the small-amplitude oscillations
in the case of aperiodic arrangement of traps, we relate our results
with those reported in the various works dealing with the Fourier
transform of these sequences, most of them in connection with X-ray
diffraction and Raman scattering in aperiodic semiconductor
superlattices, and neutron scattering in quasicrystals.
\cite{IEEE,Benoit,Godreche,Merlin2,Axel} Let us consider, in the first
place, the case of the FL, which is known to exhibit a pure point
Fourier spectrum displaying well defined Bragg peaks.\cite{Merlin2} The
Fourier intensity $|\langle 0 |{\cal H}_{\text T}| k\rangle|^{2}$ of a
FLs consists of a series of peaks located at values of the momentum
$2\pi(k/N)$ of the form $2\pi(m+n\sigma)$, where $m$ and $n$ are two
arbitrary integers and $\sigma \equiv (\sqrt{5}+1)/2$ is the golden
mean.  Each peak of the Fourier intensity leads to a large contribution
to the sum (\ref{antefinal}) and, consequently, to a well-defined
frequency in the oscillatory pattern of $p(t)$.  A simple inspection of
Fig.~\ref{fig2}(a) indicates that the pattern of $p(t)$ is dominated by
a single frequency and, in fact, the Fourier analysis of the data gives
the approximate value $\sim 3.456$ for this main frequency.  The
frequency of the particular mode $k_p$ that significantly contributes to
the oscillatory pattern is then given by
$\omega_{k_p}-\omega_{0}=3.456$, leading to $k_0/N\simeq 0.3798$, the
ratio being independent of the system size.  Therefore the mode is
located close to the top of the excitonic band ($k/N=0.5$).

\begin{figure}

\centerline{\epsfxsize=6.0truecm \epsffile{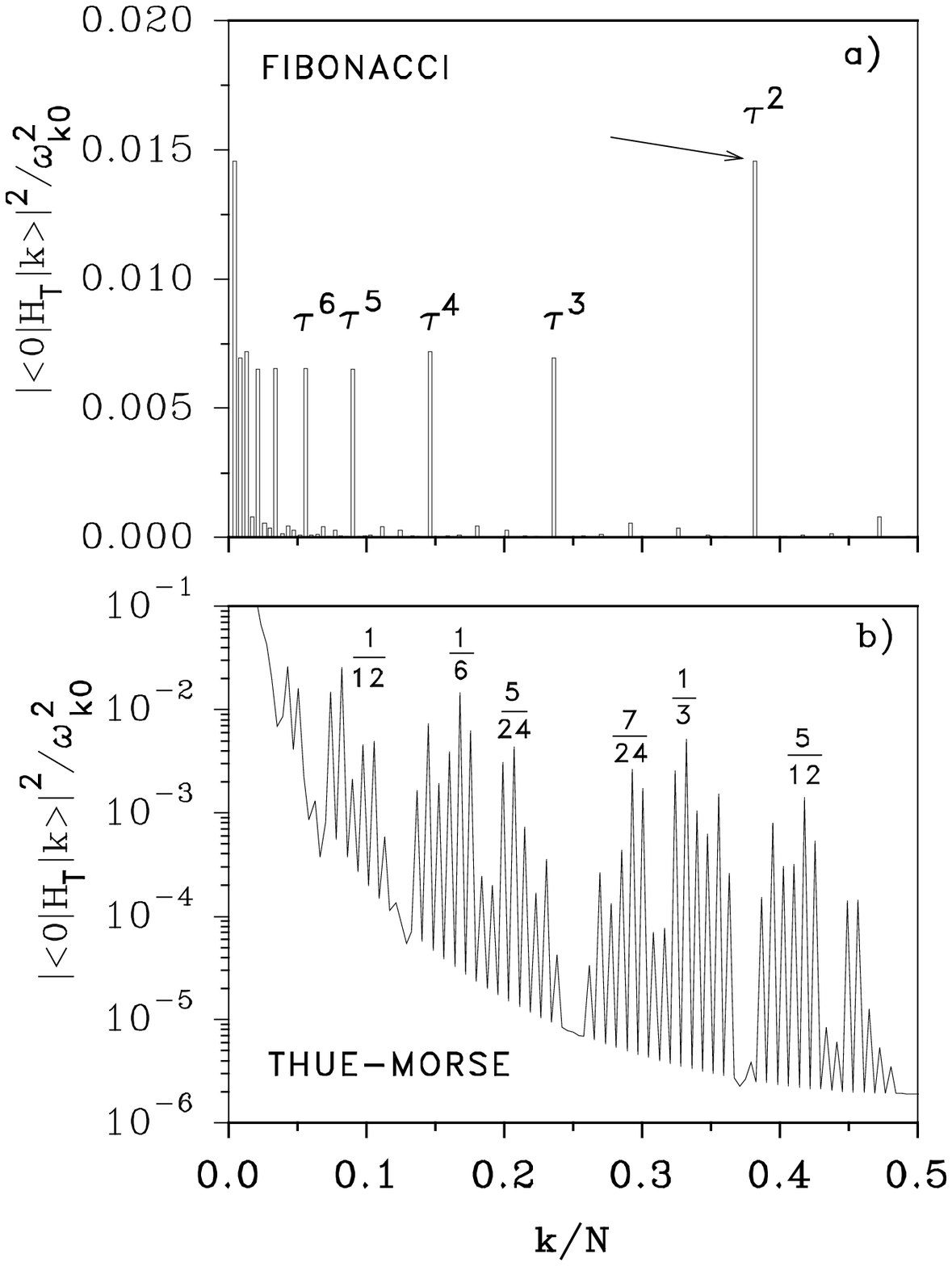}}\vspace{-1cm}

\caption{(a) Plot of $|\langle 0 |{\cal H}_{\text T}|k\rangle |^{2}/
\omega^{2}_{k0}$ as a function of the ratio $k/N$ for FLs showing the
occurrence of well-defined Bragg peaks, labeled according to successive
powers of the inverse golden mean $\tau$.  The arrow indicates the peak
responsible of the largest-frequency oscillations observed in Fig.~2.
(b) Plot of $|\langle 0|{\cal H}_{\text T} | k \rangle|^{2} /
\omega^{2}_{k0}$ as a function of the ratio $k/N$ for a TML with
$2^{8}=256$ active centers.  A dense arrangement of peaks labeled
according to the formula $(2n+1)/(3\times 2^{m})$ is clearly observed.}

\label{fig3}

\end{figure}

Figure~\ref{fig3}(a) shows $|\langle 0 |{\cal H}_{\text T}| k\rangle
|^{2}/\omega^{2}_{k0}$ as a function of the ratio $k/N$ evaluated from
expression (\ref{final}), the results being independent of the system
size.  A sequence of Bragg peaks, arranged according to successive
powers of $\tau$, are clearly observed in this plot.  The stronger one,
which is responsible of the highest-frequency oscillations, is located
at $k/N \simeq 0.38\simeq \tau^2$.  Making use of the relationships
$\tau^2+\tau=1$ and $\sigma=1+ \tau$, we find the two indices labeling
this peak to be $m=2$ and $n=-1$.  The Fourier pattern is symmetric
around the top of the excitonic band and, in fact, there exists also a
similar peak located at $k/N \simeq 0.62 \simeq 1-\tau^{2}=\tau$, whose
indices are $m=-1$ and $n=1$.  The remaining, in general weaker, peaks
also give rise to oscillations in $p(t)$ but their contribution is not
appreciable in the decay curve because of they are closer to the bottom
of the excitonic band and then lead to low-frequency oscillations.

Let us consider now the case of the TML, which is known to exhibit a
singular continuous Fourier transform, lacking well defined Bragg
peaks.\cite{Axel} For finite TMLs the Fourier intensity consists of a
dense arrangement of peaks whose number depends on the sample size.  For
a TML composed of $2^{n}$ building blocks of the same size one expects
to find $2^{n-1}$ peaks in the related Fourier transform whose positions
are located at values of the momentum $2\pi (k/N)$ of the form
$(2n+1)/(3\times 2^{m})$, where $n$ and $m$ are two arbitrary
integers.\cite{Axel} An inspection of the Fig.~\ref{fig2}(b) indicates
that, conversely to the case of FL, the pattern of the oscillatory
fluorescence decay corresponding to the TML cannot be described in terms
of a single main frequency.  In fact, the Fourier analysis of numerical
data shows a considerable number of frequencies closely packed.
Figure~\ref{fig3}(b) shows $|\langle 0 |{\cal H}_{\text T}|
k\rangle|^{2}/\omega_{k0}^2$ as a function of the ratio $k/N$ for a TML
with $N=2^{8}$ active centers.  A dense distribution of peaks is clearly
observed.  We have checked that most peaks can be labeled by means of
the formula $(2n+1)/(3\times 2^{m})$.  This we show in the figure for
the most relevant peaks.  By comparing Fig.~\ref{fig3}(b) with X-ray
results obtained for semiconducting TM superlattices, \cite{Axel} we see
that the study of the oscillations in the fluorescence decay allow us to
analyze the topological structure for a wide variety of aperiodic
systems, even if those system do not give rise to well defined Bragg
peaks in the reciprocal space.

Our results are suitable for a direct physical interpretation.
Small-amplitude oscillations of largest frequency $\omega_{k_p}-
\omega_{0}$ are caused by the coupling of two modes, namely the
lowest-lying and $k_p$, through the topology of the aperiodic
distribution of traps.  Notice that different arrangement of traps would
lead to completely different Fourier intensity $|\langle 0 |{\cal
H}_{\text T}|k\rangle|^{2}$ and $p(t)$, as it can be seen from a
comparison of Figs.~\ref{fig2}(a) and (b).  Thus, the exciton acts as a
probe of the spatial distribution of traps.  This is one of the main
results of the present work since it provides us with a useful method to
be applied in experimental situations.  To elucidate the particular
arrangement of traps in a linear lattice, first we can analyze the
fluorescence decay after $\delta$-pulse excitation.  If this decay is
stretched exponential, the arrangement of traps is completely
disordered, whereas simple exponential behavior indicates {\em absence
of disorder} corresponding to either periodic or aperiodic distribution
of trapping centers.  The slope of the average exponential decay permits
to find out the concentration of traps through the exponent value of
$P_{\text{HM}}(t)$, and then $p(t)\sim \ln [P(t)/P_{\text{HM}}(t)]$.  By
performing the Fourier analysis of $p(t)$ one can determine its related
power spectrum and then, comparing the obtained results with theoretical
predictions from (\ref{antefinal}) and (\ref{final}), to extract
relevant information about the topological order of the underlying
structure.

On the other hand, from a theoretical point of view, this work
represents a twofold extension of the mathematical treatment originally
introduced by Huber. \cite{Huber92} In the first place, instead to study
the asymptotic fluorescence decay in {\em random} Frenkel lattices we
consider the case of general {\em aperiodic} systems.  In the second
place, meanwhile Huber evaluated the dipole moment matrix elements using
unperturbed eigenvectors and perturbed eigenvalues, we show that this
approach can only explain the occurrence of long-period oscillations.
To account for short-period oscillations we must include the second term
of (\ref{pervector}), that is, perturbed eigenvectors.  This second term
is responsible for the coupling between the lowest-lying mode and the
high-lying modes through the aperiodic arrangement of traps, causing the
short-period oscillations.

\section{Conclusions}

In this paper we have carried out a detailed analysis of the
fluorescence decay in self-similar Frenkel lattices with interstitial
traps arranged according to the Fibonacci and Thue-Morse sequences.  To
this end, we have solved numerically the equation of motion of Frenkel
excitons in the time domain.  Unlike what it was previously found in
random lattices, we have observed that the fluorescence decay after
$\delta$-pulse excitation, in which all centers are initially excited,
behaves exponentially on average in the case of aperiodic systems.  We
have demonstrated that such a time decay is quite well described within
a homogeneous medium approximation, where the aperiodic distribution of
traps is replaced by a uniform distribution one with a renormalized
trapping rate.  This exponential trend is similar to that found in the
case of incoherent transport through aperiodic lattices and, in spite of
its dependence on the concentration of traps, it cannot be used to fully
characterize each particular arrangement of trapping centers.  Instead,
quantum effects reveal themselves through small-amplitude oscillations
superimposed to the average exponential decay.  Interestingly, we have
shown that the oscillatory pattern is characteristic of each particular
lattice and, consequently, it can be considered as a fingerprint that
allows us to elucidate the kind of underlying ordering present in the
distribution of traps.

We have carried out a perturbation approach to account for the
small-amplitude oscillations, finding an excellent agreement with
numerical data.  The perturbative analysis clearly indicates that the
oscillatory pattern of the fluorescence decay is directly related to the
Fourier transform of the distribution of traps and, as a consequence, it
is characteristic of the ordering present in the considered system.  In
this sense, we realize that the necessary condition for an arbitrary
system to exhibit Fourier peaks is the occurrence of long-range order in
its structure.  This condition plays a major role in our theoretical
analysis.  Since each well-defined frequency contributing to the
function $p(t)$ can be related to a distinctive peak in the Fourier
transform of the lattice, the Fourier analysis of the time-dependent
signal gives us direct information concerning the spatial distribution
of traps.  The physical origin of the fluorescence oscillations can be
traced back to the coupling between the low-lying exciton mode and
certain particular high-lying modes through the lattice topology.  This
means that excitons act as a probe of the spatial distribution of
trapping centers.  Hence the detailed analysis of time-decay of the
fluorescence, which uses excitons as a physical probe of the long-range
order, represents a powerful tool to investigate the topology of general
aperiodic systems from experiments, opening in this way a novel
experimental technique which may be tentatively referred to as {\em
laser crystallography}.

\acknowledgments

The authors thank A.\ S\'{a}nchez and V.\ Malyshev for a critical
reading of the manuscript.  This work is supported by CICYT (Spain)
through project MAT95-0325.

\end{document}